\documentclass[aps,prl,twocolumn,superscriptaddress,footinbib]{revtex4-2}

\usepackage{graphicx}  
\usepackage{subfigure}
\usepackage{dcolumn} 
\usepackage{bm}
\usepackage{amssymb}   
\usepackage{amsfonts}   
\usepackage{comment}
\usepackage[colorlinks,linkcolor=blue,anchorcolor=blue,citecolor=blue,urlcolor=blue]{hyperref}
\usepackage{xcolor}
\usepackage{soul}
\usepackage{tabularx,booktabs,ragged2e,multirow}
\newcolumntype{C}{>{\centering\arraybackslash}X}
\usepackage{multirow}
\usepackage{amsmath}
\usepackage{mathrsfs}
\usepackage{mathcomp}
\usepackage{textcomp}
\usepackage{dsfont}
\usepackage{esint}
\usepackage{braket}
\usepackage{textgreek}
\usepackage{lipsum}
\usepackage{marvosym} 
\usepackage{centernot}
\usepackage{cancel}
\usepackage{dashrule}

\begin{document}
\preprint{APS/123-QED}

\title{Dissipation Enhanced Unidirectional Transport in Topological Systems}

\author{Ming Lu}
\affiliation{Beijing Academy of Quantum Information Sciences, Beijing 100193, China}

\author{Xue-Zhu Liu}
\affiliation{Center for Advanced Quantum Studies, Department of Physics,
Beijing Normal University, Beijing 100875, China}

\author{Hailong Li}
\affiliation{International Center for Quantum Materials, School of Physics, Peking University, Beijing 100871, China}

\author{Zhi-Qiang Zhang}
\email{zhangzhiqiangphy@163.com}
\affiliation{School of Physical Science and Technology, Soochow University, Suzhou 215006, China}
\affiliation{Institute for Advanced Study, Soochow University, Suzhou 215006, China}
\author{Jie Liu}
\email{jieliuphy@xjtu.edu.cn}
\affiliation{ Department of Applied Physics, School of Science, Xian Jiaotong University, Xian 710049, China}

\author{X.C. Xie}
\affiliation{International Center for Quantum Materials, School of Physics, Peking University, Beijing 100871, China}
\affiliation{Institute for Nanoelectronic Devices and Quantum Computing, Fudan University, Shanghai 200433, China}
\affiliation{Hefei National Laboratory, Hefei 230088, China}
\date{\today}

\newcommand{\br}{{\bm r}}
\newcommand{\bk}{{\bm k}}
\newcommand{\bq}{{\bm q}}
\newcommand{\bp}{{\bm p}}
\newcommand{\bv}{{\bm v}}
\newcommand{\bmm}{{\bm m}}
\newcommand{\bA}{{\bm A}}
\newcommand{\bE}{{\bm E}}
\newcommand{\bB}{{\bm B}}
\newcommand{\bH}{{\bm H}}
\newcommand{\bd}{{\bm d}}
\newcommand{\bzero}{{\bm 0}}
\newcommand{\bOmega}{{\bm \Omega}}
\newcommand{\bsigma}{{\bm \sigma}}
\newcommand{\bJ}{{\bm J}}
\newcommand{\bL}{{\bm L}}
\newcommand{\bS}{{\bm S}}
\newcommand\dd{\mathrm{d}}
\newcommand\ii{\mathrm{i}}
\newcommand\eff{\mathrm{eff}}
\newcommand\ee{\mathrm{e}}
\newcommand\zz{\mathtt{z}}
\makeatletter
\let\newtitle\@title
\let\newauthor\@author
\def\ExtendSymbol#1#2#3#4#5{\ext@arrow 0099{\arrowfill@#1#2#3}{#4}{#5}}
\newcommand\LongEqual[2][]{\ExtendSymbol{=}{=}{=}{#1}{#2}}
\newcommand\LongArrow[2][]{\ExtendSymbol{-}{-}{\rightarrow}{#1}{#2}}
\newcommand{\cev}[1]{\reflectbox{\ensuremath{\vec{\reflectbox{\ensuremath{#1}}}}}}
\newcommand{\red}[1]{\textcolor{red}{#1}} 
\newcommand{\blue}[1]{\textcolor{blue}{#1}} 
\newcommand{\green}[1]{\textcolor{orange}{#1}} 
\newcommand{\mytitle}[1]{\textcolor{orange}{\textit{#1}}}
\newcommand{\mycomment}[1]{} 
\newcommand{\note}[1]{ \textbf{\color{blue}#1}}
\newcommand{\warn}[1]{ \textbf{\color{red}#1}}

\makeatother


\begin{abstract}
Dissipation is a common occurrence in real-world systems and is generally considered to be detrimental to transport. In this study, we examine the transport properties of a narrow quantum anomalous Hall system with dissipation applied on one edge. When the Fermi level resides within the hybridization gap, we find that while transport is suppressed on one edge, it is significantly enhanced on the other. We reveal that this enhancement arises from dissipation-induced gap closure, which is deeply rooted in the point gap topology of the system, resulting in a reduction of the decaying coefficient.\mycomment{By adjusting the dissipation amplitude, we identify two distinct exceptional transitions where both the real and imaginary parts of the band gap close at the transition points.} When the dissipation is very large, we find that the low-energy physics is nearly indistinguishable from a narrower system, whose dissipation amplitude is inversely proportional to that of the original one. To get more physical intuition, we demonstrate that the low-energy physics can be well captured by a pair of coupled counter-propagating chiral edge states, one of which has a modified group velocity and an effective dissipation. We also briefly discuss the possible experimental realizations of this enhanced unidirectional transport.
\end{abstract}



\maketitle

\section{Introduction}
Non-Hermitian (NH) physics has drawn intensive research interests recently\cite{Bender1998, Bender2002, ElGanainy2018, Ashida2020}. Many unique features have been uncovered which has no Hermitian counterparts, including exceptional degeneracies\cite{Dembowski2001, Dembowski2004, Berry2004,Heiss2012, Wiersig2014, Zhen2015, Xu2017,Chen2017, Hodaei2017, Cerjan2018,Cerjan2019,Zhang2019, Miri2019,Wiersig2020,Wiersig2020a, Bergholtz2021, Ding2022}, extended topological classifications\cite{Shen2018,Gong2018, Kawabata2019, Kawabata2019a, Torres2019, Ghatak2019}, and NH skin effect (NHSE)\cite{Zhang2022,Okuma2023, Lin2023}. Among them, NHSE is of particular importance. It leads to the failure of conventional bulk-boundary correspondence and motivates the crucial concepts such as generalized Brillouin zone and point gap topology\cite{Lee2016, Alvarez2018, Yao2018, Xiong2018,Lee2019, Yao2018a, Kunst2018, Yokomizo2019, Kawabata2020, Yang2020, Zhang2023, Liu2023}. Macroscopic states are aggregated near the boundary of a system in NHSE, originating from the point gap topology that exists only in NH Hamiltonians\cite{Zhang2020,Okuma2020}. Meanwhile, the protection of the real line gap may also lead to the localization of states near the system boundary, namely the topologically protected edge states\cite{Hasan2010,Qi2011,Chiu2016, Bansil2016, Schindler2018,Xie2021}. A growing trend is the study of the interplay between these two kinds of topology and their corresponding boundary phenomena\cite{Lee2019a,Li2020, Zou2021, Zhang2021, Li2022,Zhu2022, Zhu2023, Ma2023, Schindler2023, Nakamura2023, Luo2019, Zhang2019a, Edvardsson2019, Liu2019, Okugawa2020,Kawabata2020a,Lu2021, Fu2021}. As a nontrivial example, when the NHSE is manifested only on the topological edge modes but not on the bulk, higher order skin effect becomes possible. It shows higher dimensional localization without the requirement of high-order topology. To realize this, several studies \mycomment{(methods? proposals? works?)} have been put forward\cite{Lee2019a,Li2020, Zou2021, Zhang2021, Li2022,Zhu2022, Zhu2023, Ma2023, Schindler2023, Nakamura2023}. A rather universal and convenient way is proposed in Ref.~\cite{Nakamura2023}, in which a constant dissipation term is applied on one of the $(d-1)$-dimensional surfaces of a $d$-dimensional topological material. As we know, the topological edge states play an essential role in the transport properties of quantum materials, whereas the study of the interplay between dissipation and the topological edge states is still in its infancy\cite{Wang2019, Marguerite2019, Sato2019, Shavit2020,Fedorova2020,Huang2020,Lin2022}. Previous works have shown that dissipation at the end of a one-dimensional Majarana chain can stabilize Majarana zero mode, improving its potential for topological quantum computing\cite{SanJose2016, Avila2019, Liu2022}. Yet, transport studies on the effects of dissipation on topological edge states in higher dimensional topological materials are still rare\mycomment{elusive?}.

In this work, we investigate the transport properties of a narrow quantum anomalous Hall system with dissipation applied upon one of its edges. When the Fermi level resides inside the hybridization gap due to finite size effect\cite{Zhou2008}, we find that with appropriate dissipation strength, transport can be significantly enhanced on the opposite edge. The reason for this enhancement is due to the closing of the real energy gap by dissipation, which results in a reduction of the decay rate for the current density. As we increase the dissipation amplitude from zero, we find that the real part of the band gap first closes and then reopens, while the imaginary part first reopens and then closes, resulting in two separate exceptional transitions. When the dissipation becomes very large, we find the low-energy physics is almost identical to that of a narrower system, with very small dissipation amplitude that is inversely proportional to the original one. To gain more physical insight, we show that low-energy physics can be effectively described by a pair of counter-propagating chiral edge states that are coupled together. One of these states has a modified group velocity and an effective dissipation. Experimentally, we briefly discuss some possible realizations of this enhanced unidirectional transport phenomenon.


\begin{figure*}[t]
\includegraphics[width=18cm]{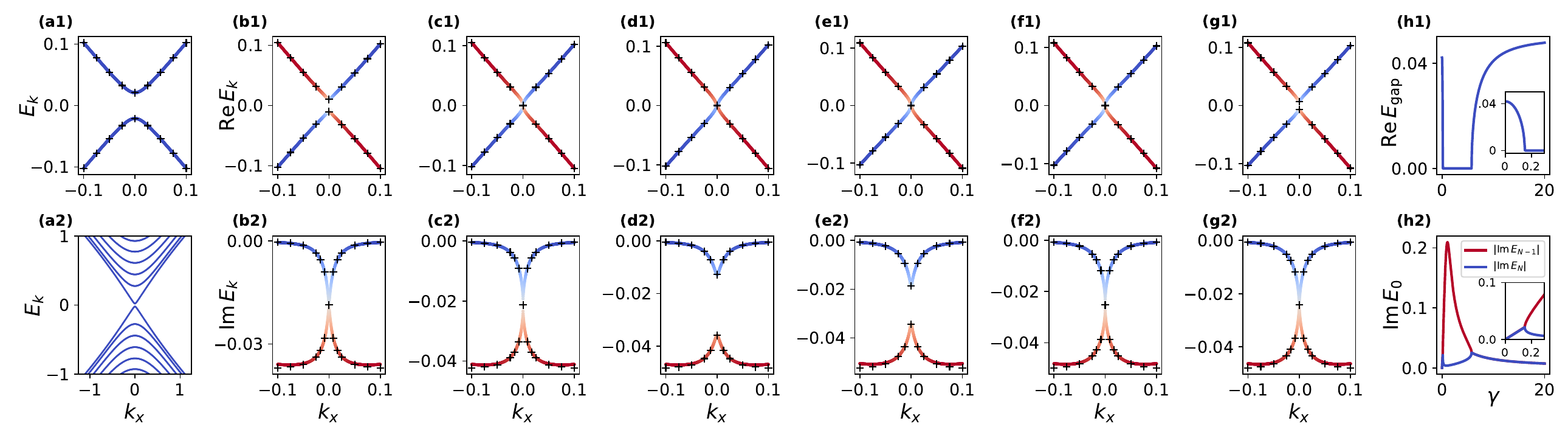}
\caption{Band structures of the ribbon with width $N_y=16$ and $m=-1.85$ and different dissipation magnitudes, where in (a) $\gamma=0$; (b) $\gamma=0.13$; (c) $\gamma=\gamma_{c1}=0.14894$; (d) $\gamma=0.17$; (e) $\gamma=5.5$; (f) $\gamma=\gamma_{c2}=5.76628$; (g) $\gamma=6.0$. The continuously changing color represents the magnitude of the imaginary part, and the black crosses represent the fit from Eq.~\ref{eq:band_structure}. The gap of the real energy band (h1) and imaginary parts of two edge states with $k_x=0$ (h2) as a function of $\gamma$. The insets are the enlarged view for the small $\gamma$ values.}\label{Fig1}
\end{figure*}

\section{Model system with edge dissipation}\label{sec2}
We consider a quantum anomalous Hall model proposed by Qi, Wu and Zhang (QWZ) \cite{Qi2006}
\begin{align}
    H = \sin k_x \sigma_x + \sin k_y \sigma_y + (m+\cos k_x+\cos k_y)\sigma_z \,.
\end{align}
This model is topological as long as $|m|<2$. When a constant dissipation is added at the $y=0$ edge, the Hamiltonian for a ribbon with $N_y$ sites in the $y$ direction reads
\begin{align}\label{Eq2}
    H(k_x)&=\sum_{n=0}^{N_y-1}[\sin k_x \sigma_x + (m+\cos k_x) \sigma_z] c_n^\dagger c_n  \nonumber \\ 
    &  + \sum_{n=0}^{N_y-2} \left[(\frac{i}{2}\sigma_y + \frac{1}{2}\sigma_z) c_{n+1}^\dagger c_{n} + \mathrm{h.c.}\right] -i\gamma\sigma_0 c_0^\dagger c_0 \,.
\end{align}
We fix the parameters $m=-1.85$ and $N_y=16$ in the following unless otherwise stated. 

As shown in Fig.~\ref{Fig1}(a), in the absence of dissipation, the hybridization of the counter-propagating chiral edge states occurs due to finite system width, opening up a small band gap around zero energy. As the dissipation is applied upon the $y=0$ edge, the gap decreases and the energy bands develop a gapless imaginary part, as shown in Fig.~\ref{Fig1}(b). When $\gamma = \gamma_{c1}$ [Fig.~\ref{Fig1}(c)], both the real and the imaginary band gaps close and the system has an exceptional degeneracy. Increasing the dissipation further, the imaginary part of the bands develops a gap and the real part remains gapless, see Fig.~\ref{Fig1}(d) and Fig.~\ref{Fig1}(e). This behavior holds until arriving at the second exceptional point $\gamma=\gamma_{c2}$ [Fig.~\ref{Fig1}(f)], where both the real and imaginary parts of band gap become gapless again. Beyond $\gamma_{c2}$, the real part of the gap opens and the imaginary part keeps gapless, as shown in Fig.~\ref{Fig1}(g). From Fig.~\ref{Fig1}(b) to Fig.~\ref{Fig1}(g), we can clearly see that the magnitude of the imaginary parts of the right-moving states are smaller than that of the left-moving ones, except at the rare places where exceptional degeneracies occur. This is simply because the dissipation is added at $y=0$ sites, which is far from the right-moving chiral edge states. The dissipation is transferred from the bottom to the top by the hybridization effect, giving the right-moving chiral edge state an effective dissipation. Therefore, the effective dissipation for the right-moving state is unavoidable given the existence of the coupling of the counter-propagating edge state and is expected to rapidly decrease with the increasing of the system width. 

The evolution of the energy gap as a function of dissipation strength can be more clearly seen in Fig.~\ref{Fig1}(h). As shown in Fig.~\ref{Fig1}(h1), the real part of the energy gap rapidly decreases to zero from $\gamma = 0$ to $\gamma_{c1}$, keeping at zero between $\gamma_{c1}$ and $\gamma_{c2}$, and then increases from zero for $\gamma > \gamma_{c2}$ and gradually saturates to a new gap value, which is slightly larger than the gap without dissipation. The closing of the hybridization gap is a result of the large mismatch of the imaginary part of the energies between the counter-propagating chiral edge states\cite{Zhou2008, Wang2015, Shafiei2022}, as well as the point gap topology dictated by the extended Nielsen-Ninomiya theorem\cite{Bessho2021}. For large enough dissipation, the system behaves like a finite width ribbon with one slice of sites narrower and with very small dissipation, leading to the increased coupling between edge states and thus a larger hybridization gap (see also Sec.~\ref{sec5}). The imaginary parts of chiral edge states at $k_x=0$ are shown in Fig.~\ref{Fig1}(h2), which bifurcate at $\gamma_{c1}$ and then merge at $\gamma_{c2}$, and finally slowly decrease towards zero for very large dissipation. The relatively small imaginary part for the left-moving edge states when $\gamma$ is very small or very large indicates the effective dissipation is small in these regimes; while the imaginary part of it is large between the two critical points, indicating the effective dissipation is large.

\begin{figure}[t]
\includegraphics[width=8.6cm]{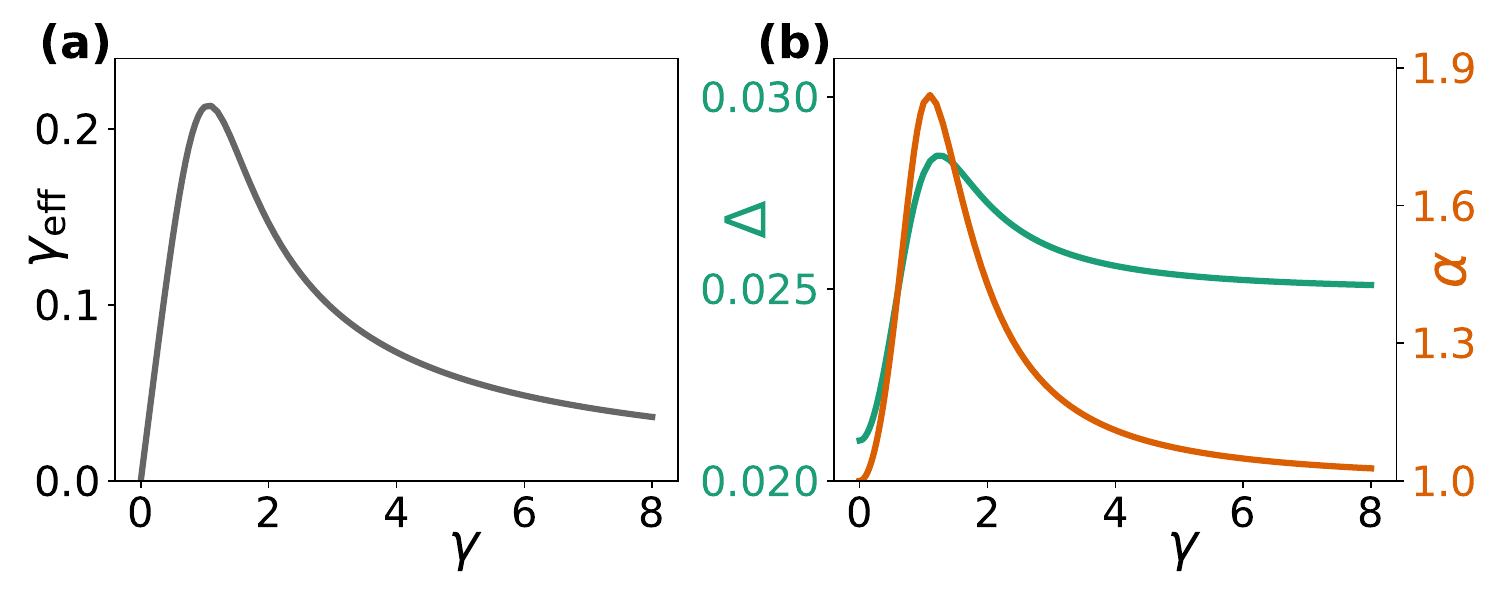}
\caption{Fitted parameters for the continuum model Eq.~(\ref{Eq:contiuum model}). (a) $\gamma_{\eff}$ as a function of $\gamma$. (b) $\Delta$ (green) and $\alpha$ (orange) as a function of $\gamma$.}\label{Fig2}
\end{figure}

The low-energy physics can be well captured by a continuum model, which describes two coupled counter-propagating chiral edge states with effective dissipation applied on one of them. The effective model reads
\begin{equation}\label{Eq:contiuum model}
    H_{\mathrm{eff}}(k_x)=\begin{pmatrix}
v k_x & \Delta \\
\Delta & -\alpha v k_x -i\gamma_{\eff} 
\end{pmatrix} \,,
\end{equation}
where $v$ and $\alpha v$ are the group velocities of the right and left-moving edge states, respectively; and $\Delta$ is the coupling strength of them. $\gamma_{\mathrm{eff}}$ is the effective dissipation, which differs from the bare dissipation $\gamma$ applied on the lattice edge. The band structure can be easily obtained
\begin{small}
\begin{equation}\label{eq:band_structure}
    E_{\pm}(k_x) = - \frac{\ii\gamma_{\eff}+(\alpha-1) v k_x}{2}  \pm \sqrt{\left(\frac{(\alpha+1)v k_x+\ii\gamma_{\eff}}{2}\right)^2+\Delta^2} \,.
\end{equation}
\end{small}Note that $\Delta$, $\alpha$ and $\gamma_\eff$ depend on $\gamma$, and can be determined in the following sense. At $k_x=0$, $E_\pm = -\ii\gamma_{\eff}/2 \pm \sqrt{\Delta^2-\gamma_\eff^2/4}$, giving $\gamma_\eff = -\mathrm{Im}\,(E_+ + E_-)$ and $\Delta = \sqrt{(E_+ - E_-)^2 + \gamma_\eff^2/4}$. We fix $v=1$ from the inspection of the Hermitian limit and determine $\alpha$ by the slope of $E_+(k_x)+E_-(k_x)$. The dependence of these parameters on $\gamma$ is given in Fig.~\ref{Fig2}. We can see the coupling strength $\Delta$ first increases, reaching the peak value at about $\gamma\approx 1$, and then approaches a new larger value. The effective dissipation also peaks at about $\gamma\approx 1$ and gradually decreases to zero for larger $\gamma$. Notably, we have equal group velocities for the small and large $\gamma$, while they become rather unbalanced in the intermediate range (see also Appendix \ref{AP:Unbalanced}). With these parameter values, the band structures calculated from the lattice model can be well fitted by Eq.~(\ref{eq:band_structure}), as shown by the black crosses in Fig.~\ref{Fig1}.

\section{Dissipation enhanced unidirectional transport}
We have shown that edge dissipation can effectively manipulate the low-energy band structures, which motivates us to study its impact on transport properties. We consider a two terminal device with two semi-infinite metal leads connected to a finite width system described by Eq.~(\ref{Eq2}). The two terminal linear conductance is calculated by $G_{pq}(E)=\mathrm{Tr}[\Gamma_p G^r \Gamma_q G^a]$, where $\Gamma_p=\ii(\Sigma_p^r-(\Sigma_p^r)^\dagger)$ is the line width function  of lead $p$ ($=L, R$), and $\Sigma_p^r$ is the self-energy due to the coupling of the lead $p$ with the central region\cite{Datta1995, Jiang2009}. The retarded Green's function is defined as $G^r(E)=(E-H_c-\sum_p\Sigma_p^r)^{-1}$, with $H_c$ represent the finite size lattice Hamiltonian of the central region. We can also calculate the local current between neighboring site $i$ and $j$ using the formula\cite{Jiang2009, Jauho1994}
\begin{equation}
    J_{ij} = \frac{2e^2 V}{h}\mathrm{Im}\sum_{\alpha,\beta} H_{i\alpha, j\beta} G^p_{j\beta, i\alpha},
\end{equation}
where the sum is over the internal degree of freedom $\alpha,\beta$ of the lattice site. $G^p = G^r\Gamma_p G^a$ is the electron correlation function and $V$ is the voltage difference between the left and right leads.

\begin{figure}[t]
\includegraphics[width=8.6cm]{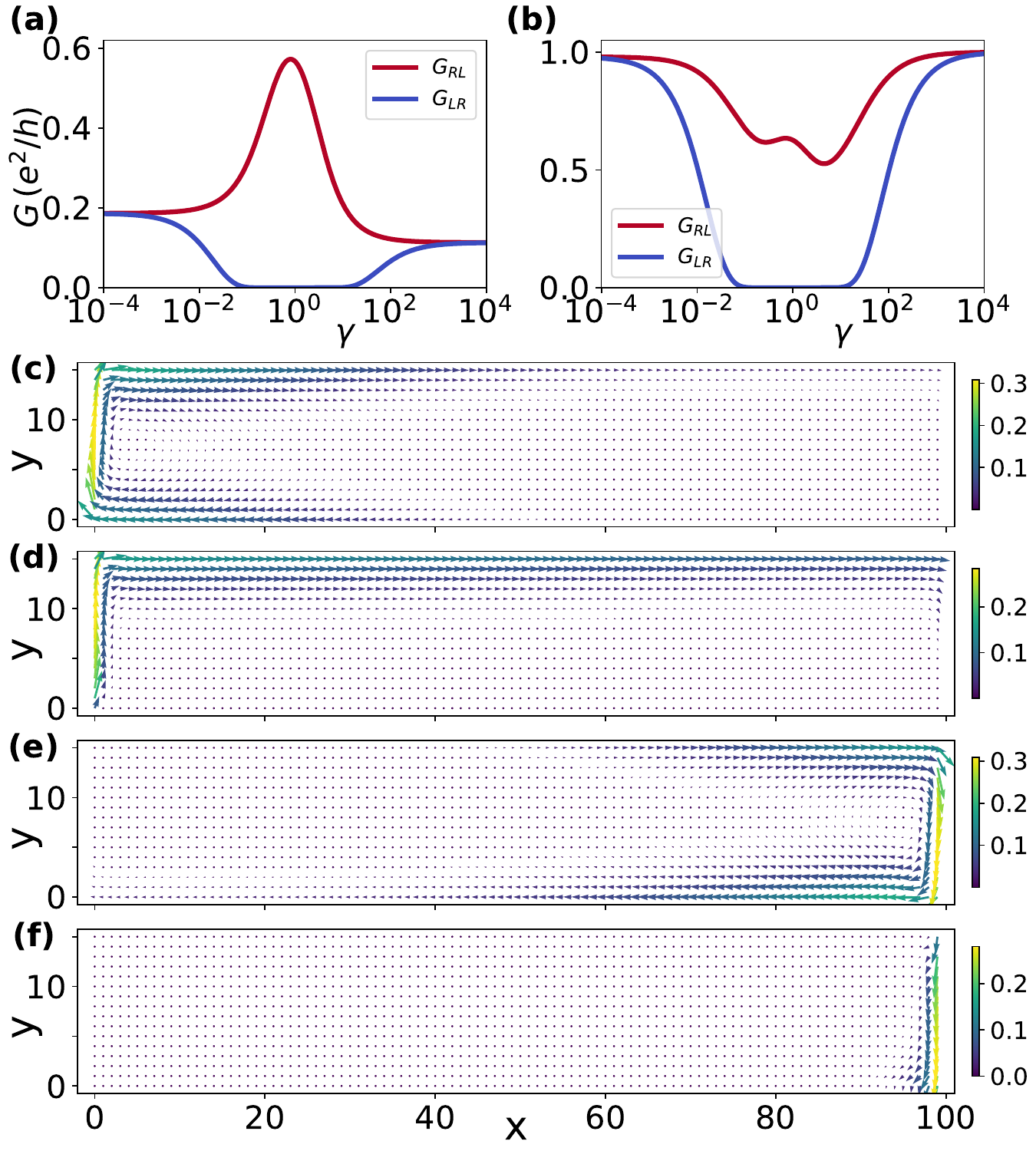}
\caption{Two terminal conductance $G_{RL}$ and $G_{LR}$ for $E_f=0.01$ (a) which is inside the hybridization gap and $E_f=0.045$ (b) which is outside the hybridization gap as a function of $\gamma$. Current density distribution for $\gamma=0$ (c) and $\gamma=0.8$ (d) when $V_L>V_R$, and for $\gamma=0$ (e) and $\gamma=0.8$ (f) when $V_L<V_R$.}\label{Fig3}
\end{figure}

In Fig.~\ref{Fig3}, we show the linear conductance and local current distribution for a system with $N_x=100$. When the Fermi level is in the gap, as shown in Fig.~\ref{Fig3}(a) with $E_f=0.01$, we can clearly see that the conductance from left to right and from right to left are distinct for a wide range of dissipation values. For some parameter regimes, the unidirectional transport is found where $G_{LR}$ is almost zero while $G_{RL}$ is finite, which is a manifestation of NHSE and point gap topology in the context of quantum transport\cite{Song2019, Yi2020}. Quite amazingly, dissipation can even enhance the transport from left to right, where $G_{RL}$ rises as dissipation increases from zero and reaches the peak when $\gamma\approx 0.8$. In the Hermitian limit where there is no dissipation, we have $G_{LR}=G_{RL}\neq 0$. The nonzero value of the conductance is the result of finite length which allow part of the current tunneling through, see also Fig.~\ref{Fig3}(c) and Fig.~\ref{Fig3}(e). When $\gamma$ is very large, we can see $G_{RL}$ and $G_{LR}$ become identical but smaller than that when $\gamma=0$, indicating for very large dissipation, the system behaves like a Hermitian system with a slightly larger gap value. When Fermi energy $E_f=0.045$ is out of the gap as shown in Fig.~\ref{Fig3}(b), we also find the dissipation induced unidirectional transport for the intermediate dissipation regime. For very large or very small dissipation, the system is in the Hermitian limit and the transmission probability is large and tends to unity.

The enhanced transport can be more clearly seen in the local current distribution in Fig.~\ref{Fig3}(c) with $\gamma=0$ and Fig.~\ref{Fig3}(d) with $\gamma=0.8$, where the left lead has the higher voltage. As shown in in Fig.~\ref{Fig3}(c), most of the current coming from left lead on top of the sample is reflected back to the left-moving channel on the bottom, and only a small fraction is managed to transmit through the sample to the right lead. On the other hand, when the dissipation is added, the left-moving channel in the bottom has a strong dissipation and thus is unable to carry the otherwise reflected current. Therefore, the reflection is suppressed and the transmission is enhanced, as clearly shown in Fig.~\ref{Fig3}(d). The unidirectional transport is also clearly seen by comparison between Fig.~\ref{Fig3}(f) and Fig.~\ref{Fig3}(d). As mentioned, the large effective dissipation destroys the bottom left-moving edge state, so that there is no channel to carry the current from right to left.


\section{suppression of the decay rate by dissipation}
\begin{figure}[t]
\includegraphics[width=8.8cm]{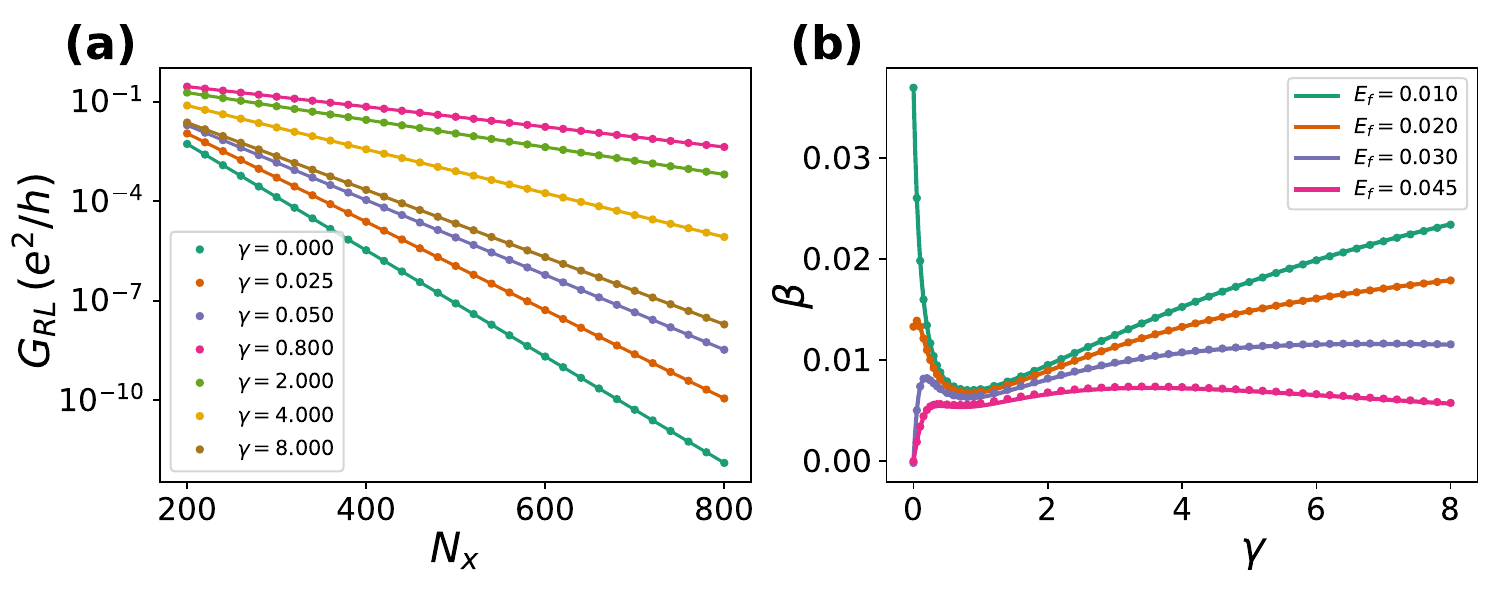}
\caption{(a) $G_{RL}$ as a function of $N_x$ for different $\gamma$'s when $E_f=0.01$. The dots are the numerical results from transport calculations and the lines are the exponential fit $A e^{-\beta N_x}$. (b) Decay rate $\beta$ as a function of $\gamma$ for different Fermi energies. The dots are the exponential fitted values and the lines are the analytical results from Eq.~(\ref{eq:decay rate}).}\label{Fig4}
\end{figure}

The enhancement of the transport by dissipation can also be explained by the decreasing of the imaginary part of the wave vector. As shown in Fig.~\ref{Fig4}(a), we calculate $G_{RL}$ by changing the system length $N_x$ for several dissipation amplitudes with fixed Fermi energy $E_f=0.01$. As clearly shown, $G_{RL}$ decays exponentially with respect to the system length. The decay rate varies non-monotonically when increasing $\gamma$: for $\gamma$ smaller than $0.8$, the slope decreases; while for $\gamma$ larger than $0.8$, the slope increases. Fig.~\ref{Fig4}(b) shows the decay rate for different Fermi energies with respect to $\gamma$, which agrees with the observation in Fig.~\ref{Fig4}(a). For $E_f$ residing inside the hybridization gap, the decay rate starts from a nonzero value and decreases rapidly with $\gamma$ and then slowly increases again. On the other hand, when $E_f$ is outside the gap, the decay rate is zero at the beginning, rapidly increasing when $\gamma$ is small, and then varies non-monotonically with the dissipation amplitude.

The decay rate can be extracted from the continuum model Eq.~(\ref{Eq:contiuum model}). To do this, we make the substitution $k_x\to-\ii \partial_x$ and assume the right-moving state with energy $E$ has the form $\psi_E(x) = e^{\ii q x} \psi_E(0)$. We can then get the expression for the wave vector $q$ by solving the Sh\"ordinger equation $H_\eff(x)\psi_E(x)=E\psi_E(x)$. We obtain
\begin{equation}\label{eq:decay rate}
    q_{\pm} = \frac{1}{2\alpha v}\left[ (\alpha-1)E-\ii\gamma_\eff\pm\sqrt{[(\alpha+1)E+\ii\gamma_\eff]^2-4\alpha\Delta^2} \right].
\end{equation}
Clearly, $q_+$ is the corresponding wave vector for the state moving along the positive $x$ direction. The current carried by state $\psi_E(x)$ is $J(x)=\frac{e\hbar}{m}\mathrm{Im}[\psi^*(x)\partial_x\psi(x)]\propto e^{-2\mathrm{Im}\,q_+ x}$. Therefore, the decay rate of the $G_{RL}$ with respect to the system length is simply $\beta(E,\gamma)=2\mathrm{Im}\,q_+(E,\gamma)$, where we have implicitly use the fact that parameters such as $\gamma_\eff, \alpha$ and $\Delta$ are functions of $\gamma$. We plot the decay rate in Fig.~\ref{Fig4}(b) using the analytical expression Eq.~(\ref{eq:decay rate}), which nicely agrees with the data points from the linear conductance calculations. From the above analysis, we can see that the enhancement of the transport from left to right is a result of dissipation enabled decreasing of the imaginary wave vector of the right-moving edge states.

\section{The large dissipation limit}\label{sec5}

\begin{figure}[t]
\includegraphics[width=8.8cm]{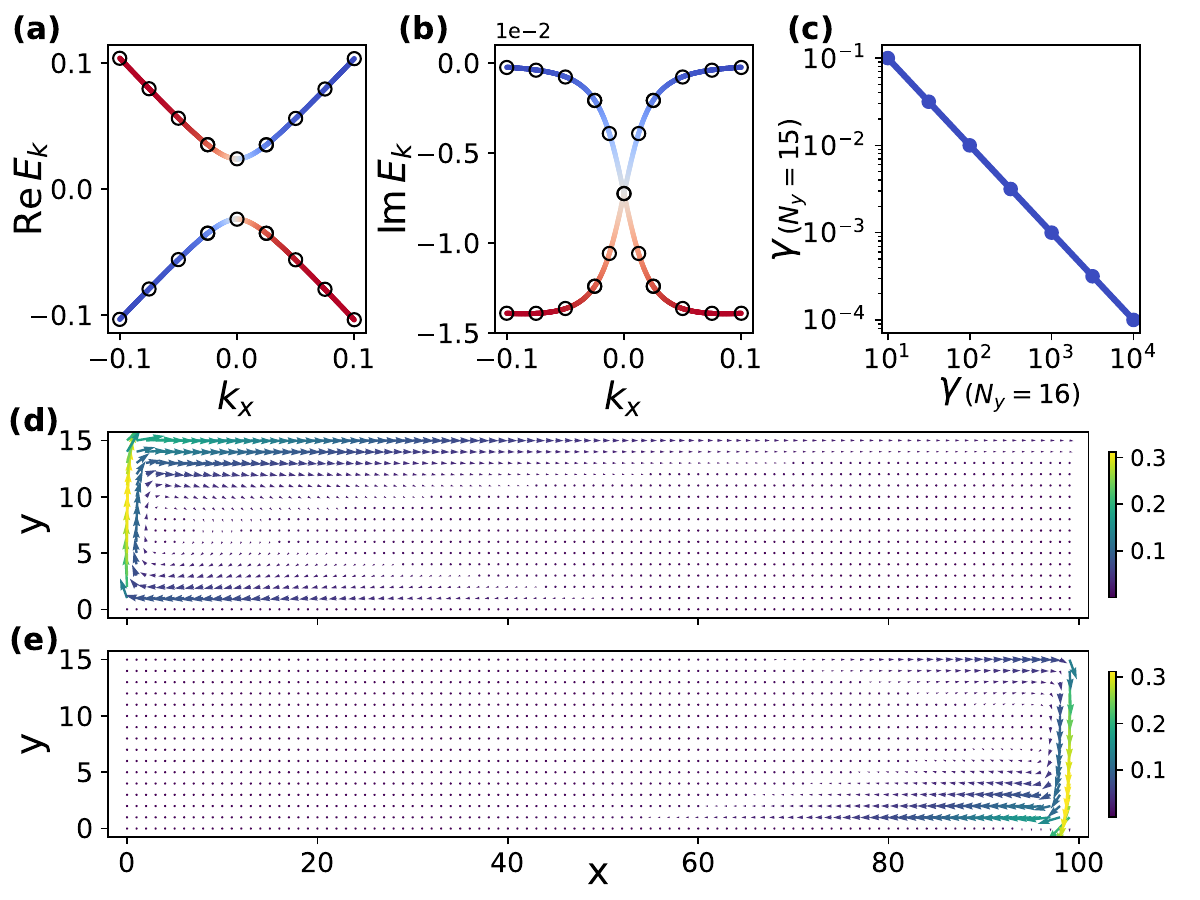}
\caption{Real (a) and imaginary (b) part of the low-energy bands. The continuous changing color lines (black circles) represent the lattice model with $N_y=16$ and $\gamma=20$ ($N_y=15$ and $\gamma=0.05$). (c) The dissipation magnitudes for system with $N_y=16$ and the corresponding values for system with $N_y=15$. The current density distribution for system with $\gamma=20$ when $V_L> V_R$ (d) and $V_L<V_R$ (e).}\label{Fig5}
\end{figure}

In this section, we explicitly show that in the large dissipation limit, the low-energy physics of the original system is almost identical to a system that is one slice narrower with very small dissipation applied on the edge. In Fig.~\ref{Fig5}(a) and Fig.~\ref{Fig5}(b), we plot the band structure of the system with $N_y=16$ and $\gamma=20$ using the continuously changing color line. Upon it, we also plot several eigenvalues of the system with $N_y=15$ and $\gamma=0.05$ marked as black circles. All other parameters are identical. As we can see, both the real part and the imaginary part of the bands of these two systems matches quite accurately. To see the equivalency more directly, we plot the current distribution for the system with Fermi energy $E_f=0.01$ and $\gamma=20$. As clearly shown in Fig.~\ref{Fig5}(d) and Fig.~\ref{Fig5}(e), the current density avoids the sites with very large dissipation and flows one site above it, behaving like a system with one slice narrower with stronger edge state coupling. The effective small dissipation for this narrower system is manifested from the asymmetric transport, which shows that the edge current from right to left decays slightly faster than that from left to right. In Fig.~\ref{Fig5}(c), we compared more dissipation amplitudes for $N_y=15$ and $N_y=16$ systems. Quite amazingly, it shows that the effective dissipation for the narrower system is precisely inverse proportional to that of the original system. In Appendix \ref{Ap:correspondence}, we give a formal proof for this observation. Thus as previously mentioned, the system will be Hermitian-like with one slice narrower in the very large dissipation limit.

\section{Summary and Discussion}

We find that edge dissipation can significantly enhance the unidirectional transport of the narrow quantum anomalous Hall system when the Fermi level resides inside the hybridization gap. We reveal that the enhancement is due to the gap closing by the dissipation, which in turn decreases the decay rate of the current density. When varying the dissipation strength, we find two exceptional transitions, where both the real and imaginary part of the band gap close. As the dissipation becomes very large, we show that the low-energy physics for the system is almost identical to a one slice narrower system whose dissipation is inversely proportional small. Using an effective continuum model, which describes two coupled counter-propagating chiral edge states with different group velocities and dissipation, we are able to capture most of the low-energy physics from the lattice calculations. Experimentally, the quantum anomalous Hall effect has been realized in several physical systems where the dissipation can be implemented in principle, including real quantum materials\cite{Chang2013, Chang2023}, cold atoms systems\cite{Jotzu2014, Liang2023}, acoustic systems\cite{Xiao2015}, \mycomment{and electric circuit systems\cite{Hofmann2019, Ezawa2019},} and photonic crystal systems\cite{Wang2009,Lu2014,Chen2022}. In Appendix \ref{app_metal}, we explicitly show that this enhanced unidirectional transport can be realized in a quantum anomalous Hall material with a coupled metallic lead acts as the source of dissipation. At last, we believe that the edge dissipation effect discussed in this work is quite general and can potentially be extended to other topological quantum systems such as quantum spin Hall insulators and topological superconductors.

\section{Acknowledgement}
We are grateful for the valuable discussions with Qing Yan, Ming Gong, Haiwen Liu and Hua Jiang. This work is supported by the Innovation Program for Quantum Science and Technology (Grants No.~2021ZD0302400) and National Natural Science Foundation of China (Grants No.~12204044, No.~12304052, No.~11974271, No.~ 92265103). Hailong Li is also funded bv China Post-doctoral Science Foundation (Grant No. BX20220005).

\appendix
\section{\MakeUppercase{Unbalanced Group Velocities in the Intermediate Dissipation Regime}}\label{AP:Unbalanced}

\begin{figure}[h]
\includegraphics[width=8.8cm]{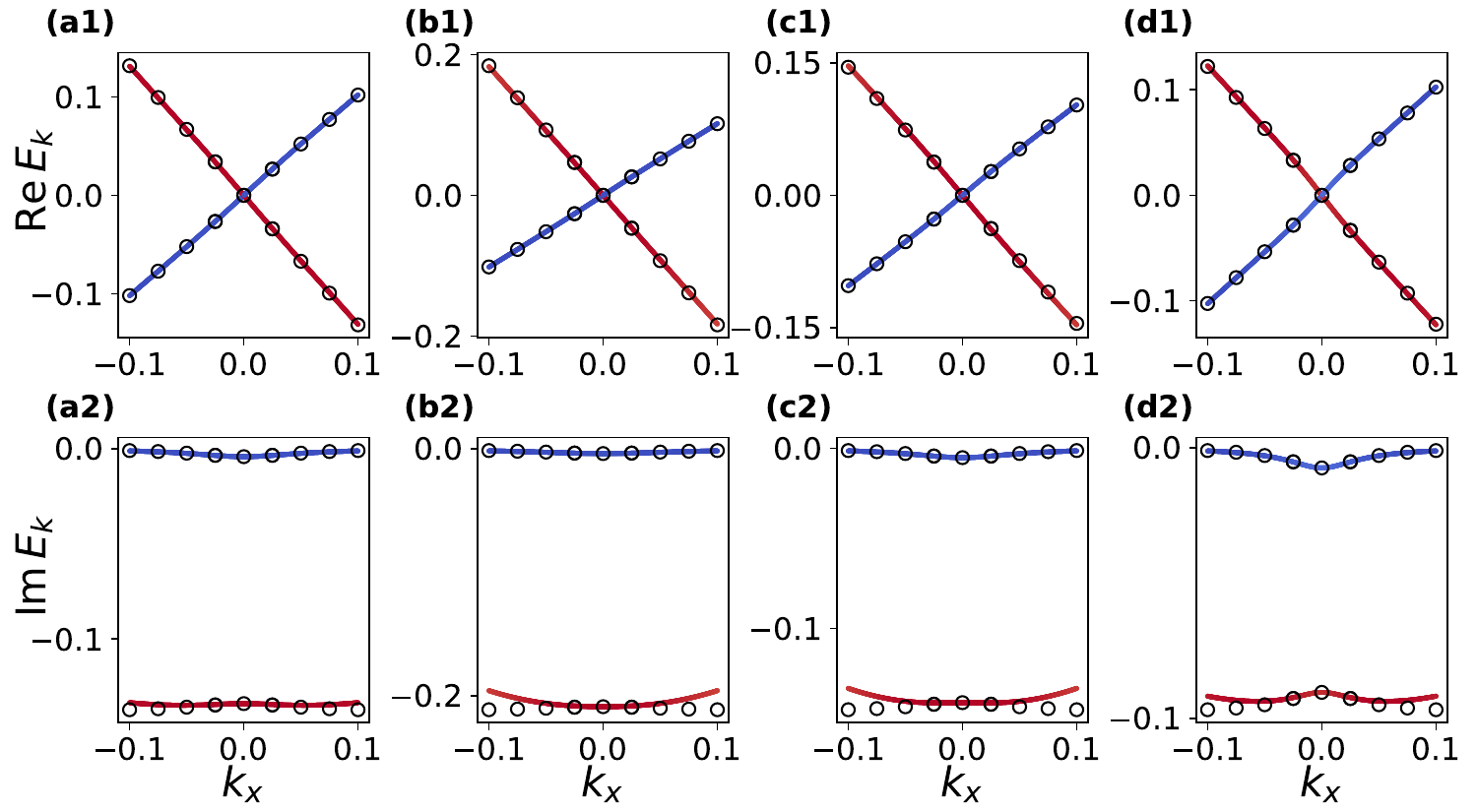}
\caption{Band structures for intermediate $\gamma$ values, where in (a) $\gamma=0.5$, (b) $\gamma=1$, (c) $\gamma=2$, and (d) $\gamma=3$. The lines are from numerical calculations of lattice model, and the black circles are from the continuum model result Eq.~(\ref{eq:band_structure}).}\label{Fig6}
\end{figure}

In this part, we show the band structures for the intermediate $\gamma$ regime. As previously shown in Fig.~\ref{Fig2}, the effective dissipation $\gamma_\eff$, the coupling strength $\Delta$, and the group velocity $\alpha$ for the left-moving state varies dramatically in this range. In Fig.~\ref{Fig6}, we plot the low-energy band structures and the corresponding continuum model fits of four different $\gamma$ values. As clearly shown, the group velocity for the left-moving edge state deviates from that of the right-moving state quite apparently, making the real part of the band structures looks like a tilted Dirac cone. The continuum model fit (black circles) for the real part of the band structure is quite accurate. However, the lower imaginary band structure is not well fitted by the continuum model. Nevertheless, the upper imaginary band is well fitted, which represents the imaginary part of the right-moving edge states. Therefore, the wave vectors of the right-moving states can still be well captured by the continuum model, as we have already shown in Fig.~\ref{Fig4}(b). The reason for the group velocity change by dissipation is not clear at this moment, and the low-energy model for a better fit of the lower imaginary energy band also need further investigation in the future.

\section{\MakeUppercase{Correspondence in the Large Dissipation Limit}}\label{Ap:correspondence}

In this part, we show that the low-energy spectrum of $H(N_y)$ and $\tilde{H}(N_y-1)$ are nearly identical when $\gamma$ is very large, where
\begin{equation}
\label{equ:A1:eq1}
\begin{split}
&H(N_y) = \left( {\begin{array}{*{20}{c}}
{{h_{00}} - i\gamma {\sigma _0}}&{{h_{01}}}&0& \cdots &0\\
{{h_{10}}}&{{h_{00}}}&{{h_{01}}}& \ddots & \vdots \\
0& \ddots & \ddots & \ddots &0\\
 \vdots & \ddots &{{h_{10}}}&{{h_{00}}}&{{h_{01}}}\\
0& \cdots &0&{{h_{10}}}&{{h_{00}}}
\end{array}} \right),  \\[5mm]
\tilde H &(N_y - 1) = \left( {\begin{array}{*{20}{c}}
{{h_{00}} - i\frac{1}{\gamma }{\sigma _0}}&{{h_{01}}}&0& \cdots &0\\
{{h_{10}}}&{{h_{00}}}&{{h_{01}}}& \ddots & \vdots \\
0& \ddots & \ddots & \ddots &0\\
 \vdots & \ddots &{{h_{10}}}&{{h_{00}}}&{{h_{01}}}\\
0& \cdots &0&{{h_{10}}}&{{h_{00}}}
\end{array}} \right).
\end{split}
\end{equation}
In the above, ${h_{00}} = \sin k_x{\sigma_x} + (m + \cos k_x){\sigma_z},\ {h_{10}} = \frac{1}{2}{\sigma _z} + \frac{i}{2}{\sigma_x},\ {\rm{ }}{h_{01}} = h_{10}^\dagger $. The Sch\"odinger equation for $H(N_y)$ is $H(N_y )\psi  = {E_{\min}}\psi$, where $E_{\min}$ and $\psi  = {\left( {\begin{array}{*{20}{c}}
{{\psi _1},}&{{\psi _2},}&{ \cdots ,}&{{\psi _{N_y}}}
\end{array}} \right)^T}$ are the energy and wave vector for the edge state, respectively. Explicitly,
\begin{align}
\left({{h_{00}}-i\gamma {\sigma_0}} \right){\psi _1} + {h_{01}}{\psi _2}&={E_{\min }}{\psi _{\rm{1}}}, \label{equ:A1:eq1} \\
{h_{10}}{\psi _1}{\rm{ + }}{h_{0{\rm{0}}}}{\psi _2}{\rm{ + }}{h_{01}}{\psi _3} &= {E_{\min }}{\psi _{\rm{2}}}, \label{equ:A1:eq2} \\
&\vdots& \\
{h_{10}}{\psi _{Ny - 1}}{\rm{ + }}{h_{0{\rm{0}}}}{\psi _{Ny}} &= {E_{\min }}{\psi _{N_y}}.
\end{align}
In the large dissipation limit, 
\begin{equation}
\label{equ:A1:eq13}
{\psi _{\rm{1}}} = (E_{\min}-h_{00}+i\gamma\sigma_0)^{-1}h_{01}\psi_2 \approx  - i\frac{1}{\gamma }{h_{01}}{\psi _2}.
\end{equation}
Substituting Eq.(\ref{equ:A1:eq13}) into Eq.(\ref{equ:A1:eq2}), we have
\begin{equation}
\left( {{h_{00}} - i\frac{1}{\gamma }{h_{10}}{h_{01}}} \right){\psi _2}{\rm{ + }}{h_{01}}{\psi _3} = {E_{\min }}{\psi _{\rm{2}}}.
\end{equation}
Therefore, the eigen-equation $H(N_y) \psi  = {E_{\min }}\psi$  transforms to $\bar H(N_y - 1)\bar\psi  = {E_{\min }}\bar \psi$, where
\begin{equation}
\bar H(Ny - 1) = \left( {\begin{array}{*{20}{c}}
{{h_{00}} - i\frac{1}{\gamma }{h_{10}}{h_{01}}}&{{h_{01}}}&0& \cdots &0\\
{{h_{10}}}&{{h_{00}}}&{{h_{01}}}& \ddots & \vdots \\
0& \ddots & \ddots & \ddots &0\\
 \vdots & \ddots &{{h_{10}}}&{{h_{00}}}&{{h_{01}}}\\
0& \cdots &0&{{h_{10}}}&{{h_{00}}}
\end{array}} \right)
\end{equation}
and $\bar\psi  = {\left( {\begin{array}{*{20}{c}}
{{\psi _2},}&{{\psi _3},}&{ \cdots ,}&{{\psi _{N_y}}}
\end{array}} \right)^T}$. 

Now we show the low-energy spectrum $\tilde{E}_{\min}$ of $\tilde{H}(N_y-1)$ can be approximated to $E_{\min }$ using perturbation theory. The Hamiltonian of the  $(N_y-1)$-wide QWZ ribbon without dissipation reads
\begin{equation}
{H_0} = \left( {\begin{array}{*{20}{c}}
{{h_{00}}}&{{h_{01}}}&0& \cdots &0\\
{{h_{10}}}&{{h_{00}}}&{{h_{01}}}& \ddots & \vdots \\
0& \ddots & \ddots & \ddots &0\\
 \vdots & \ddots &{{h_{10}}}&{{h_{00}}}&{{h_{01}}}\\
0& \cdots &0&{{h_{10}}}&{{h_{00}}}
\end{array}} \right),
\end{equation}
Its eigen-equation is written as ${H_0}{\psi ^{\left( 0 \right)}} = E_{\min }^{\left( 0 \right)}{\psi ^{\left( 0 \right)}}$,  where  ${\psi ^{\left( 0 \right)}} = {\left( {\begin{array}{*{20}{c}}
{\psi _1^{\left( 0 \right)},}&{\psi _2^{\left( 0 \right)},}&{ \cdots ,}&{\psi _{N_y-1}^{\left( 0 \right)}}
\end{array}} \right)^T}$.
To the first order perturbation, we have
\begin{align}
 E_{\min } &= E_{\min }^{\left( 0 \right)} - i\frac{1}{\gamma }\psi _1^{\left( 0 \right)\dagger }{h_{10}}{h_{01}}\psi _1^{\left( 0 \right)}, \label{equ:A1:eq17}
 \\
\tilde E_{\min } &= E_{\min }^{\left( 0 \right)} - i\frac{1}{\gamma }\ \psi _1^{\left( 0 \right)\dagger }{\sigma _0}\psi _1^{\left( 0 \right)}. \label{equ:A1:eq19}
\end{align}
where $\psi _1^{\left( 0 \right)}$ satisfies
\begin{equation}
    h_{00}\psi _1^{\left( 0 \right)} + h_{01}\psi _2^{\left( 0 \right)} = E_{\min }^{\left( 0 \right)} \psi _1^{\left( 0 \right)}.
\end{equation}
By a unitary transformation, we have
\begin{equation}\label{eq:eq_for_phi}
    \tilde{h}_{00} \phi_{1} + \sigma_{-} \phi_{2} = E_{\min }^{\left( 0 \right)} \phi_{1},
\end{equation}
where $\sigma_- = \begin{pmatrix}0&0\\1&0\end{pmatrix}$, ${{\tilde h}_{00}} = (m + \cos kx){\sigma _x} - \sin k_x{\sigma _z}$ and $\psi_{1, 2}^{(0)} = u_x \phi_{1, 2}$ with $u_x=\begin{pmatrix}1&1\\-1&1\end{pmatrix}/\sqrt{2}$. 
From Eq.~(\ref{eq:eq_for_phi}), $\phi_1$ can be written as
\begin{equation}
{\phi _1} = \left( {\begin{array}{*{20}{c}}
1\\
{ - \frac{{\left( {E_{\min }^{\left( 0 \right)} + \sin k_x} \right)}}{M}}
\end{array}} \right){\phi _{11}},
\end{equation}
where $M=-(m+\cos k_x)$. Accordingly,
\begin{align}
E_{\min } &= E_{\min }^{\left( 0 \right)} - i\frac{1}{\gamma }{\left| {{\phi _{11}}} \right|^2},\\ 
\tilde E_{\min } &= E_{\min }^{\left( 0 \right)} - i\frac{1}{\gamma }\left[ {{{\left( {\frac{{E_{\min }^{\left( 0 \right)} + \sin k_x}}{M}} \right)}^2} + 1} \right]{\left| {{\phi _{11}}} \right|^2}.
\end{align}
For small $k_x$, $M \gg E_{\min }^{\left( 0 \right)} + \sin k_x$, therefore we have $\tilde E_{\min } \approx  E_{\min }$.

\section{\MakeUppercase{Metallic Lead as the Source of Dissipation}} \label{app_metal}
\begin{figure}[t]
\includegraphics[width=8.8cm]{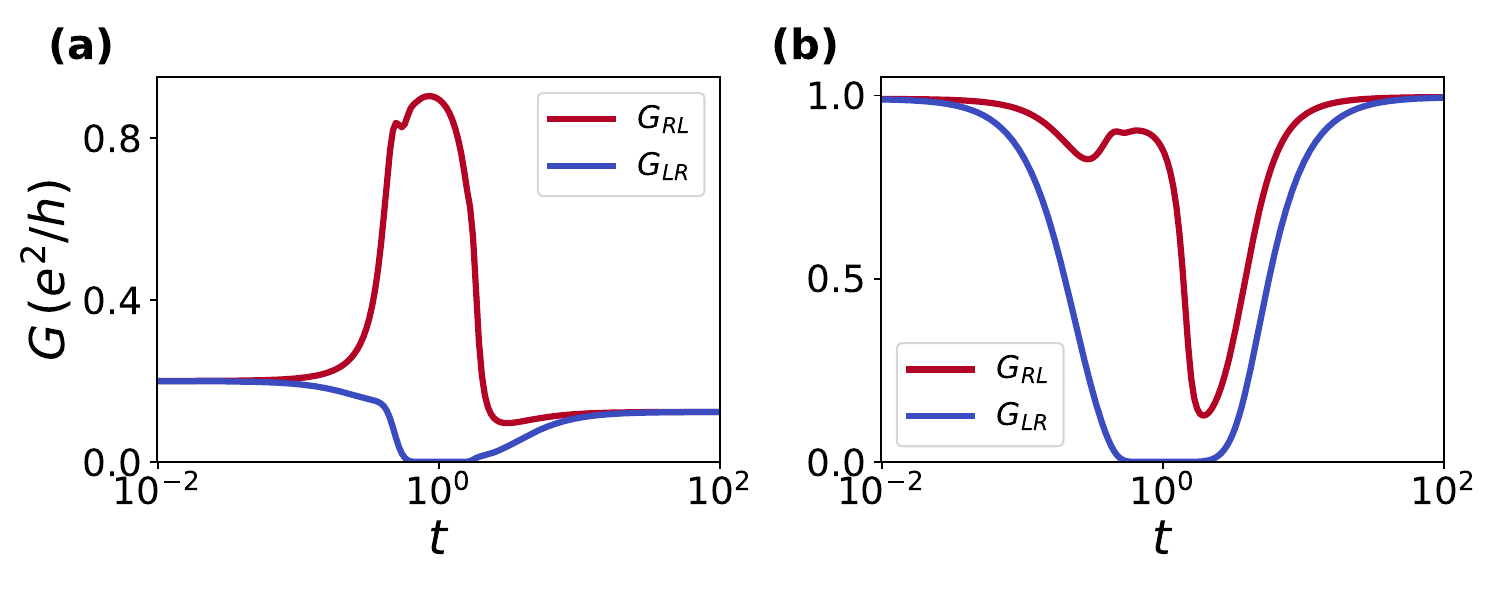}
\caption{Three terminal linear conductance $G_{RL}$ and $G_{LR}$ for $E_f=0.01$ (a) and $E_f=0.045$ (b) as a function of coupling strength $t$ with the third metallic lead.}\label{Fig7}
\end{figure}

 In the following, we explicitly show that the aforementioned dissipation enhanced transport can be realized for a quantum anomalous hall material with a metallic lead connected to its edge. The lead act as the source of dissipation, into which the electrons from the edge states can flow. From the aspect of Green's function, the coupling of the metallic lead with the central region provides the later a self-energy. This self-energy is nonzero only on the edge sites and its magnitude can be experimentally tuned by the coupling strength, which in principle can be adjusted by applying gate voltage on the coupling region. In Fig.~\ref{Fig7}, we show the linear conductance $G_{RL}$ and $G_{LR}$ as a function of the coupling strength $t$. The dissipation enhanced unidirectional transport and the recovery of the hermiticity in the large coupling limit are observed, which resembles the features shown in Fig.~\ref{Fig3}. To obtain the $G_{RL}$ in experiment, we set the voltage of left lead $V_L>0$, and connect right lead and bottom lead to the ground. $G_{RL}$ is calculated as the ratio between the current flow into the right lead $I_R$ and the voltage of left lead $V_L$. Similarly, $G_{LR}$ can be obtained by $I_L/V_R$ when left and bottom leads are grounded.

\let\oldaddcontentsline\addcontentsline
\renewcommand{\addcontentsline}[3]{}



\bibliography{references.bib}  

\end{document}